\documentclass[12pt]{article}
\usepackage{amsmath}

\begin{document}

\title{Fluctuations and control in the Vlasov-Poisson equation}
\author{Ricardo Lima\thanks{%
Centre de Physique Th\'{e}orique, CNRS Luminy, case 907, F-13288 Marseille
Cedex 9, France; lima@cpt.univ-mrs.fr} and R. Vilela Mendes\thanks{%
Centro de Fus\~{a}o Nuclear - EURATOM/IST Association, Instituto Superior
T\'{e}cnico, Av. Rovisco Pais 1, 1049-001 Lisboa, Portugal} \thanks{%
CMAF, Complexo Interdisciplinar, Universidade de Lisboa, Av. Gama Pinto, 2 -
1649-003 Lisboa (Portugal); http://label2.ist.utl.pt/vilela/} \thanks{%
Corresponding author, e-mail: vilela@cii.fc.ul.pt, vilela@cpt.univ-mrs.fr}}
\date{}
\maketitle

\begin{abstract}
In this paper we study the fluctuation spectrum of a linearized
Vlasov-Poisson equation in the presence of a small external electric field.
Conditions for the control of the linear fluctuations by an external
electric field are established.
\end{abstract}

\section{\textbf{Introduction}}

In the past, the fluctuation spectrum of charged fluids was studied either
by the BBGKY hierarchy derived from the Liouville or Klimontovich equations,
with some sort of closure approximation, or by direct approximations to the
N-body partition function or by models of dressed test particles, etc. (see
reviews in \cite{Oberman} \cite{Krommes}).

Alternatively, by linearizing the Vlasov equation about a stable solution
and diagonalizing the Hamiltonian, a method has been developed \cite
{Morrison} that uses the eigenvalues associated to the continuous spectrum
and a canonical partition function to compute correlation functions. Here
this approach will also be followed to study the control of the
fluctuations. For simplicity we will consider the one-space dimensional case.

A Vlasov-Poisson equation with an external electrical field control term is
considered. Following the method developed by Morrison \cite{M2} we use an
integral transform to solve the linearized equation. With a view to
applications to more general kinetic equations (gyrokinetic, etc.) we also
discuss in the appendix a generalization of Morrison's integral transform.

Control of the Vlasov-Poisson equation 
\[
\partial _{t}f+v\cdot \bigtriangledown _{x}f+\bigtriangledown _{x}\phi \cdot
\bigtriangledown _{v}f=C\left( t,x,v\right) 
\]
in a periodic domain 
\[
\left( t,x,v\right) \in [0,T]\times T^{n}\times R^{n} 
\]
by means of an interior control located in a spatial subdomain has been
discussed by Glass \cite{Glass}. Conditions for controllability between two
small distribution distributions $f_{0}$ and $f_{1}$ were established.
However, to steer $f_{0}$ to $f_{1}$ a control $C\left( t,x,v\right) $ that
depends on the velocities is required and it is not clear how such a control
could be implemented in practice. Therefore, we have restricted ourselves to
the more realistic situation of a (small) controlling external electric
field. In addition we concentrate on the problem of the damping of the small
oscillations around an equilibrium distribution.

In Sect.2 the linearized Vlasov-Poisson equation with control is solved by
an integral transform and in Sect.3 two controlling problems are studied,
namely the control of the total energy of the fluctuations by a constant
electric field and the dynamical damping of the fluctuating modes by a
time-dependent electric field.

\section{The linearized equation with control}

Consider a Vlasov-Poisson system in $1+1$ dimensions 
\begin{equation}
\frac{\partial f}{\partial t}+v\frac{\partial f}{\partial x}-\frac{e}{m}%
\left( \frac{\partial \Phi _{f}}{\partial x}-E^{c}\left( x,t\right) \right) 
\frac{\partial f}{\partial v}=0  \label{2.1}
\end{equation}
\begin{equation}
\frac{\partial ^{2}}{\partial x^{2}}\Phi _{f}=-e\int f\left( v\right)
dv+\rho _{B}  \label{2.2}
\end{equation}
with an external (control) electric field $E^{c}\left( x,t\right) $ and a
background charge density $\rho _{B}\left( x\right) $ chosen in a such a way
that the total charge vanishes. Consider now the linearization about a
homogeneous solution. 
\begin{equation}
f\left( x,v,t\right) =f^{(0)}\left( v\right) +\delta f\left( x,v,t\right)
\label{2.3}
\end{equation}
\[
\Phi _{f}=\Phi _{f}^{(0)}+\delta \Phi _{f} 
\]
with 
\[
\bigtriangleup \Phi _{f}^{(0)}=-e\int f^{(0)}\left( v\right) dv+\rho _{B} 
\]
such that 
\[
\bigtriangleup \Phi _{f}^{(0)}=0=\bigtriangledown \Phi _{f}^{(0)} 
\]
Then 
\[
\frac{\partial }{\partial t}f^{(0)}\left( v\right) =0 
\]
and $f^{(0)}\left( v\right) $ is indeed a homogeneous static equilibrium.
The linearized equation is 
\begin{equation}
\frac{\partial \delta f}{\partial t}+v\frac{\partial \delta f}{\partial x}-%
\frac{e}{m}\frac{\partial \delta \Phi _{f}}{\partial x}\frac{\partial f^{(0)}%
}{\partial v}+\frac{e}{m}E^{c}\left( x,t\right) \frac{\partial f^{(0)}}{%
\partial v}=0  \label{2.4}
\end{equation}
where we have assumed that 
\begin{equation}
E^{c}\left( x,t\right) =O\left( \delta f\right)  \label{2.5}
\end{equation}
that is, $E^{c}\left( x,t\right) $ is a small external (control) electric
field\footnote{%
that is, the control electric field is of the same order as the
fluctuations, not of order $\int \delta f\left( v\right) dv$, which would
lead to a trivial control situation}.

Fourier transforming all the perturbations 
\begin{equation}
\delta f\left( x,v,t\right) =\sum f_{k}\left( v,t\right) e^{ikx}
\label{2.6a}
\end{equation}
\begin{equation}
\delta \Phi \left( x,t\right) =\sum \phi _{k}\left( t\right) e^{ikx}
\label{2.6b}
\end{equation}
\begin{equation}
E^{c}\left( x,t\right) =\sum E_{k}\left( t\right) e^{ikx}  \label{2.6c}
\end{equation}
leads to 
\begin{equation}
\partial _{t}f_{k}\left( v,t\right) +ikvf_{k}\left( v,t\right) -i\frac{e^{2}%
}{mk}\left( \int f_{k}\left( \mu \right) d\mu \right) f^{(0)^{^{\prime }}}+%
\frac{e}{m}E_{k}\left( t\right) f^{(0)^{^{\prime }}}=0  \label{2.7}
\end{equation}
With a view to applications to more general kinetic equations (gyrokinetic,
etc.) a more general equation is studied in the appendix. Equation (\ref{2.7}%
) is then a particular case of Eq.(\ref{A.2}) with 
\begin{equation}
g_{1}\left( v\right) =ikv,\hspace{0.2cm}g_{2}\left( v\right) =-i\frac{e^{2}}{%
mk}f^{(0)^{\prime }}\left( v\right) ,\hspace{0.2cm}g_{3}\left( v\right) =1,%
\hspace{0.2cm}C\left( v,t\right) =\frac{e}{m}E_{k}\left( t\right)
f^{(0)^{\prime }}\left( v\right)  \label{2.8}
\end{equation}
the integral transform being (as in Morrison \cite{M2}) 
\begin{equation}
G_{k}\left( u,t\right) =\left( G_{-}f_{k}\right) \left( u\right) =\left( 1-%
\frac{\pi e^{2}}{mk^{2}}\left( Hf^{(0)^{^{\prime }}}\right) \right)
f_{k}\left( u\right) +\frac{\pi e^{2}}{mk^{2}}f^{(0)^{^{\prime }}}\left(
Hf_{k}\right)  \label{2.9}
\end{equation}
with left inverse 
\begin{equation}
G_{+}\circ \left\{ \left( 1-\frac{\pi e^{2}}{mk^{2}}\left( Hf^{(0)^{^{\prime
}}}\right) \right) ^{2}+\left( \frac{\pi e^{2}}{mk^{2}}f^{(0)^{^{\prime
}}}\right) ^{2}\right\} ^{-1}  \label{2.9a}
\end{equation}
$G_{+}$ being 
\begin{equation}
\left( G_{+}f_{k}\right) \left( u\right) =\left( 1-\frac{\pi e^{2}}{mk^{2}}%
\left( Hf^{(0)^{^{\prime }}}\right) \right) f_{k}\left( u\right) -\frac{\pi
e^{2}}{mk^{2}}f^{(0)^{^{\prime }}}\left( Hf_{k}\right)  \label{2.9b}
\end{equation}

Applying the integral transform (\ref{2.9}) to Eq.(\ref{2.7}) it becomes 
\begin{equation}
\partial _{t}\left( G_{-}f\right) \left( u\right) +iku\left( G_{-}f\right)
\left( u\right) =-\frac{e}{m}E_{k}\left( t\right) f^{(0)^{^{\prime }}}\left(
u\right)  \label{2.10}
\end{equation}
with solution

\begin{equation}
G_{k}\left( u,t\right) =e^{-itku}\left( G_{k}\left( u,0\right) -\frac{e}{m}%
f^{(0)^{^{\prime }}}\left( u\right) \int_{0}^{t}E_{k}\left( \tau \right)
e^{i\tau ku}d\tau \right)  \label{2.12}
\end{equation}
Then, according to (\ref{2.9a}), the Fourier modes solution is 
\begin{eqnarray}
f_{k}\left( v,t\right) &=&\frac{\left( 1-\frac{\pi e^{2}}{mk^{2}}\left(
Hf^{(0)^{^{\prime }}}\right) \left( v\right) \right) G_{k}\left( v,t\right) 
}{\left( 1-\frac{\pi e^{2}}{mk^{2}}\left( Hf^{(0)^{^{\prime }}}\right)
\left( v\right) \right) ^{2}+\left( \frac{\pi e^{2}}{mk^{2}}f^{(0)^{^{\prime
}}}\left( v\right) \right) ^{2}}  \label{2.13} \\
&&-\frac{\pi e^{2}}{mk^{2}}f^{(0)^{^{\prime }}}\left( v\right) H\left( \frac{%
G_{k}\left( u,t\right) }{\left( 1-\frac{\pi e^{2}}{mk^{2}}\left(
Hf^{(0)^{^{\prime }}}\left( u\right) \right) \right) ^{2}+\left( \frac{\pi
e^{2}}{mk^{2}}f^{(0)^{^{\prime }}}\left( u\right) \right) ^{2}}\right)
\left( v,t\right)  \nonumber
\end{eqnarray}

\subsection{Control of the linear modes by the electric field}

Nonlinear stability of the steady states of the Vlasov-Poisson equation when
the phase-space density is a decreasing function of the particle energy or
depend on other invariants has been studied\cite{Rein1} \cite{Rein2} by the
energy-Casimir method\cite{Holm}. This means that deviations from the
steady-state will remain bounded in time.

However, as expected from the non-dissipative nature of the Vlasov equation,
the linear fluctuation modes of the uncontrolled equation are oscillatory
and, once excited by a perturbation, they will not decay. As shown by
Morrison \cite{M2} they may be used to obtain a statistical description of
the fluctuations by the construction of a partition function. Here, one
focus on the control of the fluctuations by the external electric field. Two
situations will be considered. The first considers a constant in time
electric field and tries to minimize the total energy associated to the
fluctuations. The functional to be minimized is 
\begin{equation}
F_{1}\left( E_{k}\right) =\lim_{T\rightarrow \infty }\int_{0}^{T}dtdu\left|
G_{k}\left( u,t\right) \right| ^{2}  \label{3.1}
\end{equation}
In the second situation we allow the electric field to be time-dependent and
chosen in a such a way as to introduce a damping effect in the solution (\ref
{2.12}).

For the first case ($E_{k}$ independent of time), with the solution (\ref
{2.12}) one obtains a minimum for the functional $F_{1}$ at 
\[
E_{k}=\frac{-\int \frac{f^{(0)^{^{\prime }}}\left( u\right) }{u}\left( -%
\textnormal{Im}G_{k}\left( u,0\right) +i\textnormal{Re}G_{k}\left( u,0\right) \right) du%
}{\frac{2e}{mk}\int \left( \frac{f^{(0)^{^{\prime }}}\left( u\right) }{u}%
\right) ^{2}du} 
\]
For this electrical field $F_{1}$ is 
\begin{equation}
\begin{array}{l}
F_{1\min }=\int \left( \left( \textnormal{Re}G_{k}\left( u,0\right) \right)
^{2}+\left( \textnormal{Im}G_{k}\left( u,0\right) \right) ^{2}\right) du \\ 
-\frac{1}{2\int \left( \frac{f^{(0)^{^{\prime }}}\left( u\right) }{u}\right)
^{2}du}\left( \int \left( \frac{f^{(0)^{^{\prime }}}\left( u\right) \textnormal{Re}%
G_{k}\left( u,0\right) }{u}\right) ^{2}+\left( \frac{f^{(0)^{^{\prime
}}}\left( u\right) \textnormal{Im}G_{k}\left( u,0\right) }{u}\right) ^{2}du\right)
\end{array}
\label{3.3}
\end{equation}
a smaller value as compared to the case $E_{k}=0$, which would be $%
F_{1}\left( E_{k}=0\right) =\int \left( \left( \textnormal{Re}G_{k}\left(
u,0\right) \right) ^{2}+\left( \textnormal{Im}G_{k}\left( u,0\right) \right)
^{2}\right) du$.

In the second case one allows the electric field to be time-dependent. One
aims at controlling the fluctuation modes by an electric field induced
dynamical damping. One looks for the solution of 
\begin{equation}
\lim_{t\rightarrow \infty }\left( G_{k}\left( u,0\right) -\frac{e}{m}%
f^{(0)^{^{\prime }}}\left( u\right) \int_{0}^{t}E_{k}\left( \tau \right)
e^{i\tau ku}d\tau \right) =0  \label{3.4}
\end{equation}
obtaining 
\begin{equation}
E_{k}\left( t\right) =\frac{mk}{2\pi e}\int_{-\infty }^{\infty }\frac{%
G_{k}\left( u,0\right) }{f^{(0)^{^{\prime }}}\left( u\right) }e^{-ikut}du
\label{3.5}
\end{equation}
Then with this electric field 
\begin{eqnarray}
&&G_{k}\left( u,0\right) -\frac{e}{m}f^{(0)^{^{\prime }}}\left( u\right)
\int_{0}^{t}E_{k}\left( \tau \right) e^{i\tau ku}d\tau  \label{3.6} \\
&=&G_{k}\left( u,0\right) -\frac{k}{2\pi }f^{(0)^{^{\prime }}}\left(
u\right) \int_{-\infty }^{\infty }du^{^{\prime }}\frac{G_{k}\left(
u^{^{\prime }},0\right) }{f^{(0)^{^{\prime }}}\left( u^{^{\prime }}\right) }%
\frac{e^{ik\left( u-u^{^{\prime }}\right) t}-1}{ik\left( u-u^{^{\prime
}}\right) }  \nonumber
\end{eqnarray}
and from 
\begin{equation}
\frac{k}{2\pi }\frac{e^{ik\left( u-u^{^{\prime }}\right) t}-1}{ik\left(
u-u^{^{\prime }}\right) }\underset{t\rightarrow \infty }{\longrightarrow }%
\delta \left( u-u^{^{\prime }}\right)  \label{3.7}
\end{equation}
one sees that the electric field (\ref{3.5}) induces a dynamical damping of
the fluctuation modes.

\section{Appendix. An integral transform for linearized kinetic equations}

Morrison \cite{M2} solves the linearized Vlasov-Poisson equation by a
Hilbert transform. However, for some practical applications, the linearized
kinetic equations are more complex. For example the gyrokinetic Vlasov
equation written in gyrocenter phase-space coordinates is \cite{Brizard} 
\begin{equation}
\frac{\partial f}{\partial t}+\stackrel{\bullet }{X}\bullet \bigtriangledown
f+\stackrel{\bullet }{U}\frac{\partial f}{\partial U}=0  \label{A.1}
\end{equation}
where $\stackrel{\bullet }{X}\neq U$

This is the motivation to study an equation more general than the linearized
Vlasov-Poisson (\ref{2.7}). Linearized Fourier kinetic equations are of the
type 
\begin{equation}
\frac{\partial f\left( v\right) }{\partial t}+g_{1}\left( v\right) f\left(
v\right) +g_{2}\left( v\right) \int g_{3}\left( \mu \right) f\left( \mu
\right) d\mu +C\left( v,t\right) =0  \label{A.2}
\end{equation}
with $g_{1}$ a monotone function of $v$. Let $T$ be a transform such that 
\begin{equation}
\left( Tg_{1}f\right) \left( u\right) =\int g_{3}\left( \mu \right) f\left(
\mu \right) d\mu +g_{1}\left( \mu \right) \left( Tf\right) \left( u\right)
\label{A.3}
\end{equation}
namely 
\begin{equation}
\left( Tf\right) \left( u\right) =P\int \frac{g_{3}\left( v\right) f\left(
v\right) }{g_{1}\left( v\right) -g_{1}\left( u\right) }dv  \label{A.4}
\end{equation}
Notice that, for invertible $g_{1}$, the $T-$transform may be written in
terms of the Hilbert transform 
\begin{equation}
\left( Tf\right) \left( u\right) =\pi \left( H\frac{g_{3}f}{g_{1}^{^{\prime
}}}\circ g_{1}^{-1}\right) \left( g_{1}\left( u\right) \right)  \label{A.4a}
\end{equation}
Then, one defines 
\begin{equation}
\left( G_{+}f\right) \left( u\right) =\left( 1+\left( Tg_{2}\right) \left(
u\right) \right) f\left( u\right) +g_{2}\left( u\right) \left( Tf\right)
\left( u\right)  \label{A.5}
\end{equation}
and 
\begin{equation}
\left( G_{-}f\right) \left( u\right) =\left( 1+\left( Tg_{2}\right) \left(
u\right) \right) f\left( u\right) -g_{2}\left( u\right) \left( Tf\right)
\left( u\right)  \label{A.5a}
\end{equation}
$\left\{ \left( 1+Tg_{2}\right) ^{2}+\pi ^{2}\left( \frac{g_{2}g_{3}}{%
g_{1}^{^{\prime }}}\right) ^{2}\right\} ^{-1}G_{-}$ is a left inverse of $%
G_{+}$%
\begin{equation}
\left( G_{-}G_{+}f\right) \left( u\right) =\left\{ \left( 1+Tg_{2}\right)
^{2}+\pi ^{2}\left( \frac{g_{2}g_{3}}{g_{1}^{^{\prime }}}\right)
^{2}\right\} f\left( u\right)  \label{A.5b}
\end{equation}
as may be checked using (\ref{A.4a}) and the properties of the Hilbert
transform \cite{Tricomi}.

$G_{-}$-transforming Eq.(\ref{A.2}) one obtains 
\begin{eqnarray}
\partial _{t}\left( G_{-}f\right) +g_{1}\left( u\right) \left( G_{-}f\right)
&=&-C\left( u,t\right) +g_{2}\left( u\right) T\left( C\right) \left(
u\right) -C\left( u,t\right) T\left( g_{2}\right) \left( u\right)
\label{A.6} \\
&=&\gamma \left( u,t\right)  \nonumber
\end{eqnarray}
with solution 
\begin{equation}
G_{-}\left( u,t\right) =e^{-tg_{1}\left( u\right) }\left( G_{-}\left(
u,0\right) +\int_{0}^{t}\gamma \left( u,\tau \right) e^{\tau g_{1}\left(
u\right) }d\tau \right)  \label{A.7}
\end{equation}


\begin{thebibliography}{99}
\bibitem{Oberman}  C. R. Oberman and E. A. Williams; in \textit{Handbook of
Plasma Physics} (M. N. Rosenbluth, R. Z. Sagdeev, Eds.), pp. 279-333,
North-Holland, Amsterdam 1985.

\bibitem{Krommes}  J. A. Krommes; Phys. Reports 360 (2002) 1-352.

\bibitem{Morrison}  P. J. Morrison; Phys. of Plasmas 12 (2005) 058102.

\bibitem{M2}  P. J. Morrison; \textit{Hamiltonian description of Vlasov
dynamics: Action-angle variables for the continuous spectrum, }Institute for
Fusion Studies report IFSR-866, 1999.

\bibitem{Glass}  O. Glass; J. Diff. Equations 195 (2003) 332-379.

\bibitem{Rein1}  G. Rein; Math. Methods Appl. Sci. 17 (1994) 831-844.

\bibitem{Rein2}  P. Braasch, G. Rein and J. Vukadinovic; Siam J. Appl. Math.
59 (1998) 831-844.

\bibitem{Holm}  D. D. Holm, J. E. Marsden, T. Ratiu and A. Weinstein; Phys.
Rep. 123 (1985) 1-116.

\bibitem{Brizard}  A. Brizard; Phys. of Plasmas 2 (1995) 459-471.

\bibitem{Tricomi}  F. G. Tricomi; \textit{Integral equations}, (theor. IV in
ch. 4), Interscience, New York 1957.
\end{thebibliography}
\end{document}